\documentclass[twocolumn]{aastex62}

\usepackage{hyperref}
\usepackage{amsmath}
\usepackage[utf8]{inputenc}
\received{July 28, 2022}
\revised{August 29, 2022}
\accepted{\today}
\submitjournal{ApJL}

\shorttitle{Early JWST Galaxy Structure}
\shortauthors{Ferreira et al.}

\def\casgm20{CAS-G-M$_{20}\,$}
\def\m20{M$_{20}\,$}

\usepackage{tikz}
\usetikzlibrary{shapes,arrows}
\begin{document}

\title{Panic! At the Disks: First Rest-frame Optical Observations of Galaxy Structure at $z > 3$ with JWST in the SMACS 0723 Field} 


\correspondingauthor{Leonardo Ferreira}
\email{leonardo.ferreira@nottingham.ac.uk}

\author[0000-0003-1949-7638]{Leonardo Ferreira}
\affil{Centre for Astronomy and Particle Physics, University of Nottingham,
Nottingham, UK}

\author[0000-0003-4875-6272]{Nathan Adams}
\affiliation{Jodrell Bank Centre for Astrophysics, University of Manchester, Oxford Road, Manchester UK}

\author[0000-0003-1949-7638]{Christopher J. Conselice}
\affiliation{Jodrell Bank Centre for Astrophysics, University of Manchester, Oxford Road, Manchester UK}

\author[0000-0001-6245-5121]{Elizaveta Sazonova}
\affiliation{Department of Physics and Astronomy, Johns Hopkins University, Baltimore, MD 21218, USA}

\author[0000-0003-0519-9445]{Duncan Austin}
\affiliation{Jodrell Bank Centre for Astrophysics, University of Manchester, Oxford Road, Manchester UK}

\author[0000-0002-6089-0768]{Joseph Caruana}
\affiliation{Department of Physics, University of Malta, Msida MSD 2080}
\affiliation{Institute of Space Sciences \& Astronomy, University of Malta, Msida MSD 2080}

\author[0000-0002-0056-1970]{Fabricio Ferrari}
\affiliation{Instituto de Matemática Estatística e Física, Universidade Federal do Rio Grande, Rio Grande, RS, Brazil}

\author[0000-0002-0730-0781]{Aprajita Verma}
\affiliation{Sub-department of Astrophysics, University of Oxford, Denys Wilkinson Building, Keble Road, Oxford, OX1 3RH.}

\author[0000-0002-9081-2111]{James Trussler}
\affiliation{Jodrell Bank Centre for Astrophysics, University of Manchester, Oxford Road, Manchester UK}

\author[0000-0002-8785-8979]{Tom Broadhurst}
\affiliation{Department of Physics, University of the Basque Country UPV/EHU, E-48080 Bilbao, Spain}
\affiliation{DIPC, Basque Country UPV/EHU, E-48080 San Sebastian, Spain}
\affiliation{Ikerbasque, Basque Foundation for Science, E-48011 Bilbao, Spain}

\author[0000-0001-9065-3926]{Jose Diego}
\affiliation{FCA, Instituto de Fisica de Cantabria (UC-CSIC), Av.  de Los Castros s/n, E-39005 Santander, Spain}

\author[0000-0003-1625-8009]{Brenda L.~Frye}
\affiliation{Department of Astronomy/Steward Observatory, University of Arizona, 933 N. Cherry Avenue, Tucson, AZ 85721, USA}

\author[0000-0002-2282-8795]{Massimo Pascale}
\affiliation{Department of Astronomy, University of California, 501 Campbell Hall \#3411, Berkeley, CA 94720, USA}

\author[0000-0003-3903-6935]{Stephen M.~Wilkins} %
\affiliation{Astronomy Centre, University of Sussex, Falmer, Brighton BN1 9QH, UK}
\affiliation{Institute of Space Sciences and Astronomy, University of Malta, Msida MSD 2080, Malta}

\author[0000-0001-8156-6281]{Rogier A. Windhorst}
\affiliation{School of Earth and Space Exploration, Arizona State University,
Tempe, AZ 85287-1404}

\author[0000-0002-0350-4488]{Adi Zitrin}
\affiliation{Physics Department, Ben-Gurion University of the Negev, P. O. Box 653, Be’er-Sheva, 8410501, Israel}



\begin{abstract}

We present early results regarding the morphological and structural properties of galaxies seen with the James Webb Space Telescope (JWST) at $z > 3$ in the Early Release Observations towards the SMACS 0723 cluster field.  Using JWST we investigate, for the first time, the optical morphologies of a significant number of $z > 3$ galaxies with accurate photometric redshifts in this field to determine the form of galaxy structure in the relatively early universe.  We use visual morphologies and \textsc{Morfometryka} measures  to perform quantitative morphology measurements, both parametric with light profile fitting (Sérsic indices) and non-parametric (CAS values).  Using these, we measure the relative fraction of disk, spheroidal, and peculiar galaxies at $3 < z < 8$. We discover the surprising result that at $z > 1.5$ disk galaxies dominate the overall fraction of morphologies, with a factor of $\sim 10$ relative higher number of disk galaxies than seen by the Hubble Space Telescope at these redshifts.   Our visual morphological estimates of galaxies align closely with their locations in CAS parameter space and their Sérsic indices.

\end{abstract}



\section{Introduction} \label{sec:intro}

The James Webb Space Telescope (JWST) was launched on December 25, 2021 with its first operational image released to the public on July 11, 2022 by US~President Joe Biden.  This first image is a very deep image of the RELICS cluster SMACS J0723.3-732 \citep[SMACS 0723,][]{ebeling2010, Repp2018, Coe2019}.  SMACS 0723 is massive cluster of galaxies at $z = 0.390$ which is also known to contain an extensive collection of strong gravitational arcs with a measured and modelled mass profile \citep[e.g.,][]{Golubchik2022, Pascale2022}.    As this is the first JWST image to be released for a field where many objects have existing accurate photometric redshifts, it enables us to study the morphological evolution of galaxies with the earliest JWST data available. 

Even before the release of the raw imaging, it was clear from the publicly released promotional color image that this cluster contained a collection of red and spiral galaxies that were not obviously present in the Hubble Space Telescope (HST) imaging.   These observations provide the ideal resource for a first examination of the problem of how galaxy morphology changes from HST to JWST, and how rest-frame optical morphologies appear in the un-probed region $z > 3$. To understand this question better, we have undertaken an early analysis of this Early Release Observation (ERO) data released by the Space Telescope Science Institute (STScI) on July 13, 2022 to the public.  

This paper is a first-look study of the morphological evolution of galaxies seen in the field around SMACS 0723, giving us our first redshift-based look at how galaxy structure changes with redshift up to $z=8$.  This is the first field where this analysis can be performed due to the limited available accurate photometric redshifts in other, early release observations.

Galaxy structure and morphology are one of the key aspects for understanding galaxy evolution, and will be a key measurement that JWST will make throughout its lifetime.  Following the first servicing mission to Hubble, distant galaxies started to have their structure resolved in the mid-nineties. This revealed that faint, distant galaxies appear more peculiar and irregular than local ones, and cannot be easily classified on the Hubble sequence \citep[][]{Griffiths1994,Dressler1994, Driver1995, vandenbergh1996}.  Why that is the case has remained a major topic of discussion for almost three decades.  These early observations, however, only showed that galaxies became more peculiar at fainter magnitudes, which did not necessarily correlate with further distances. 
  
When redshifts became available, at first within the Hubble Deep Field, it was clear that galaxy structures evolve strongly and systematically with redshift, such that peculiar galaxies dominate the population at $z > 2.5$ \citep[e.g.,][]{Conselice2003a, Papovich2005, Elmegreen2005, Dahlen2007, Buitrago2008, Conselice2008, Huertas2009, Buitrago2012, Mantha2018}. It is now well-established that galaxies as observed with HST become smaller and more irregular/peculiar at higher redshifts, and this has been accounted for by the merger process for a significant fraction (40-50\%) of systems \citep[][]{Conselice2014a}.

However, although HST was revolutionary, morphological evolution measurements still suffer some limitations. First among these is that due to HST's limited red wavelength coverage, we have not measured the rest-frame optical light of galaxies within the first two Gyr after the Big Bang, that is at $z > 3$.  Very few galaxies have been observed in the rest-frame optical bands at such redshifted wavelengths, and most of these utilise ground-based adaptive optics.  This is due to the fact that the F160W band on HST only probes rest-frame optical light up to $z\sim2.8$, whereas JWST permits us to obtain this information up to $z=8$ with F444W, and even beyond with MIRI.  Second, HST infrared imaging does not provide the necessary spatial resolution to resolve most high-redshift objects. Furthermore, we have found in previous observations \citep[e.g.,][]{Conselice2005, Huertas2009,Mortlock2013, Huertas2016} that the number of galaxies that are classifiable as disks or spheroids (including ellipticals) declines quickly when observing systems at higher redshifts, up to $z = 3$.  

Observations of galaxy structure and morphology at $z > 3$ do show that in the rest-frame UV, galaxies are peculiar and irregular \citep[e.g.,][]{Elmegreen2005, conseliceandA2009}. Galaxies at these redshifts are also often found to be clumpy, as seen with deep Wide Field Camera 3 (WFC3) data \citep[][]{Elmegreen2005,Oesch2010, Margalef2018, Whitney2021, Margalef2022}.  Measurements of galaxies in pairs also demonstrates that the merger rate and the fraction of galaxies in mergers at $z \sim 6$ is as high as 50\% \citep[e.g.,][]{conseliceandA2009, duncan2019}.  This implies that galaxy structure should likewise be distorted accordingly \citep[e.g.,][]{duncan2019,shibuya2022}.  At the same time, we have believed for 30 years that the Hubble sequence is established quite early at $z \sim 1$ \citep[e.g.,][]{Mortlock2013, Huertas2016}.  However, all of these conclusions are based on HST imaging, which has now been superseded in significant ways by the redder bands, higher resolution, and better sensitivity of JWST.

Thus, in this paper we explore the morphological properties of the earliest galaxies through an approach based on galaxy classification and measurement. We demonstrate that these early galaxies have a more normal morphology than expected, with classifications showing that disk galaxies are much more common than previous observations suggested \citep[e.g.,][]{Conselice2005, Huertas2009,Conselice2014a, Huertas2016,Margalef2022}.  Overall, we argue that the formation of the Hubble sequence appears to be ongoing much earlier than we had anticipated based on HST observations.  

This paper is organized as follows: in \S 2 we describe the data and our methods and outline.  In \S 3 we describe the morphological results of our study, \S 4 is a short discussion of our results, and \S 5 is an overall summary. Throughout this paper we assume a $\Lambda$ cold dark matter cosmological model with $\Omega_{\Lambda} = 0.7$, $\Omega_{\textrm{M}} = 0.3$ and $H_0 = 70$ km s$^{-1}$ Mpc$^{-1}$. All magnitudes are given in the AB system \citep{Oke1974,Oke1983}.

\section{Data Reduction and Products}

The data we use for this analysis originates from the Early Release Observations of SMACS~0723 \citep{ERO} and include observations taken with the \textit{Near Infrared Camera} \citep[NIRCam;][]{rieke05, rieke08, rieke15}. The images were obtained on June 06, 2022 (PI: Pontoppidan; Program ID 2736) in the F090W, F150W, and F200W short-wavelength (SW) bands, and F356W, F277W, and F444W long-wavelength (LW) bands. The total integration time for this target is 12.5\,hr.  Figure~1 shows the combined color image of SMACS~0723 which we created from our own reduction.

We reprocess the uncalibrated lower-level JWST data products following a slightly modified version of the JWST official pipeline. This is because the initial release of the higher-level data products have been found to contain WCS alignment issues as well as sub-optimal background subtraction. The key differences are as follows: (1) We use version 1.5.2 of the pipeline as opposed to version 1.5.3, which was the most up-to-date version at the time of writing. This is because version 1.5.3 has a significant bug in the background subtraction step that led to sub-optimal performance.\footnote{\url{https://github.com/spacetelescope/jwst/issues/6920}} (2) We apply the CEERS 1/F noise and flat field correction (Bagley et al. in prep) between stages 1 and 2 of the official pipeline. (3) We extract the \texttt{SkyMatchStep} from stage 3 and run it independently on each NIRCam frame, allowing for quicker assessment of the background subtraction performance and fine-tuning. (4) After Stage 3, we align the final science images onto a GAIA-derived WCS using \texttt{tweakreg}, part of the DrizzlePac python package\footnote{\url{https://github.com/spacetelescope/drizzlepac}}. We then pixel match the images with the use of \texttt{astropy reproject}\footnote{\url{https://reproject.readthedocs.io/en/stable/}}.
Finally, we re-align the RELICS SMACS~0723 HST imaging to the GAIA DR2 catalog due to large $~1$\arcsec~offsets, and match it with \texttt{astropy reproject} as well.  We then apply astrometric corrections to the positions of sources available in the RELICS catalogs. 

Overall, this data set allows us to probe the rest-frame optical images of galaxies out to $z = 8$. In Figure \ref{fig:rest-frame}, we show the rest-frame wavelength probed by each individual filter and how they can be combined together for up to $z=8$ optical rest-frame coverage. In addition to this, we combine our observations with HST Archival data in the WFC3 F160W band.

\begin{figure}
    \centering
    \includegraphics[width=0.45\textwidth]{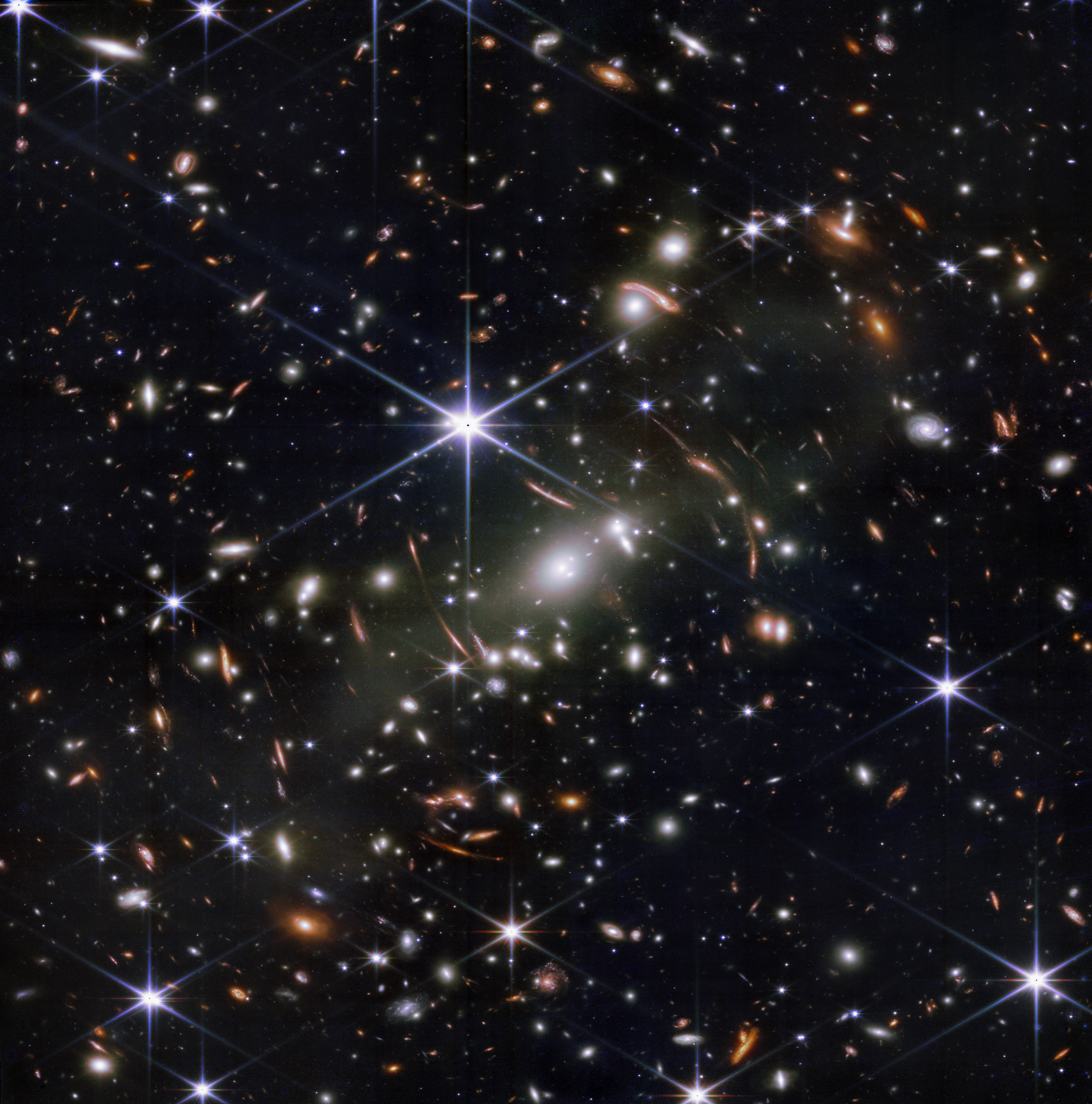}
    \caption{JWST color image of SMACS 0723 showing the overall distribution of galaxy shapes and morphologies, including the lensing arcs. This image was produced from our reduced data products via a composite of data in 6 bands: F090W, F150W, F200W, F277W, F356W, and F444W. F090W and F150W were assigned blue colours, F200W and F277W green, and F356W and F444W orange and red respectively.}
    \label{fig:my_label}
\end{figure}

We employ two different approaches to the SMACS 0723 data: first a quantitative analysis using \textsc{Morfometryka} \citep{Ferrari2015}, where we measure non-parametric morphology estimates such as concentration, asymmetry, and smoothness \citep[CAS;][]{Conselice2003}; Gini-M20 \citep{Lotz2004},  various sizes, as well as light profile fitting, which is described in detail \S \ref{sec:mfmtk}. Second, we provide simple visual classifications for all sources with $S/N > 10$ in their optical rest-frame filters, described in detail in \S \ref{sec:visual}.

\begin{figure}
    \centering
   \includegraphics[width=0.5\textwidth]{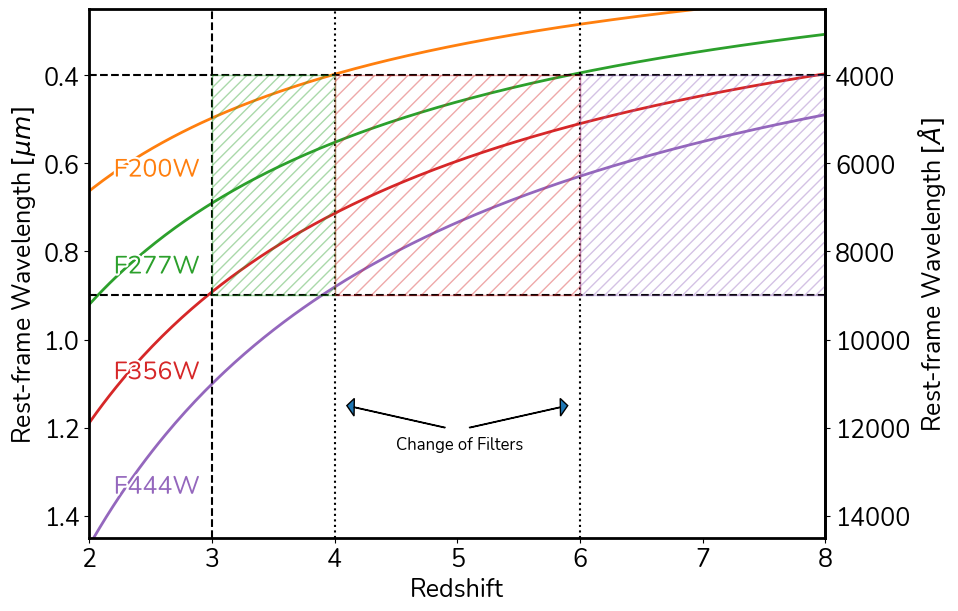}
    \caption{Rest-frame wavelength at a given redshift for the F200W, F277W, F356W, and F444W filters. The hatched regions show the areas where NIRCam filters probe the optical rest-frame for $z > 3$, with the color corresponding to the respective optimal filter for the redshift range. To follow galaxies' optical rest-frames we use F277W for $3 < z < 4$, F356W for $4 < z < 6$, and F444W for $6 < z < 8$.}
    \label{fig:rest-frame}
\end{figure}

\subsection{Photometric Redshifts}

We use photometric redshifts derived through the Bayesian photo-z code (BPz) \citep{Benitez2000, Benitez2004, DanCoe2006} by the RELICS program \citep{Salmon2020}, which used HST imaging in 7 bands for 41 clusters, including SMACS~0723, and archival \textit{Spitzer} IRAC measurements to measure photometric redshifts of galaxies up to $z=8$. The BPz code compares RELICS fluxes to 11 templates for ellipticals, spirals and starburst galaxies.  The overall method for this is described in detail in \citet[][]{Coe2019}.   

\begin{figure*}
    \centering
    \includegraphics[width=0.9\textwidth]{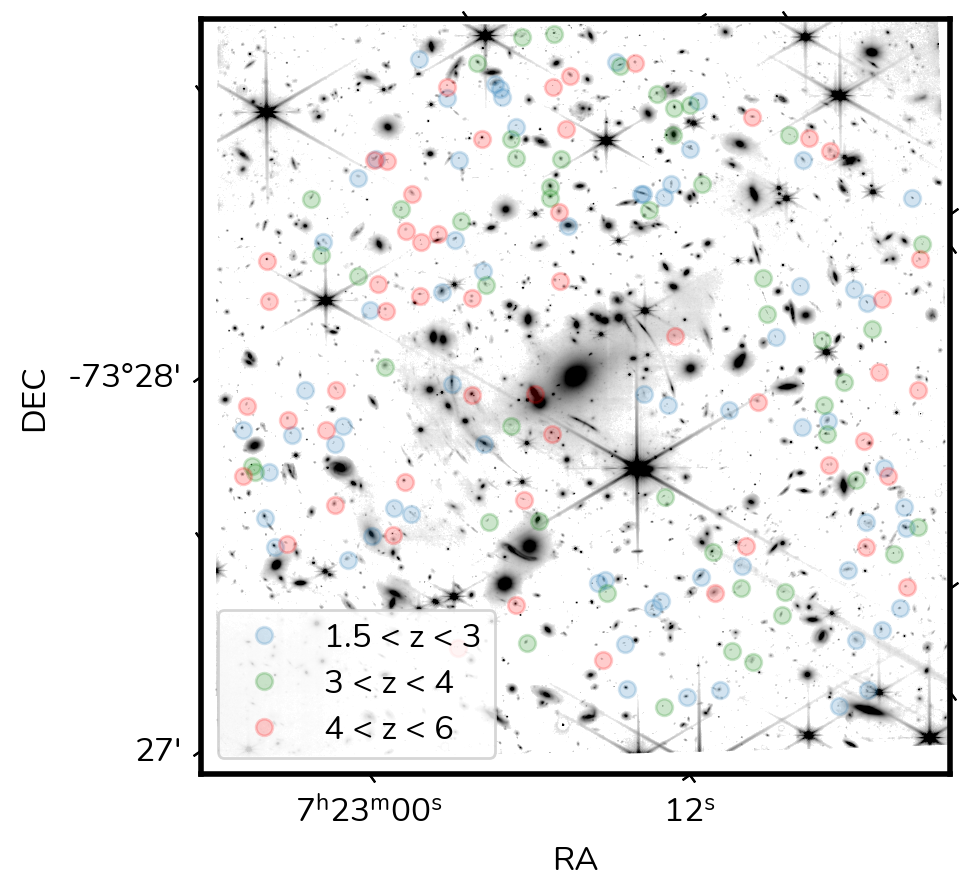}
    \caption{The distribution of redshifts across the SMACS 0723 field.  Different redshift regimes, where we use our methodology to investigate the morphological evolution, are shown as differently-colored markers.}
    \label{fig:redshifts}
\end{figure*}

For our analysis, we selected 355 galaxies with RELICS photometric redshifts from the JWST footprint of the SMACS~0723 field, restricting our sample to $1.5 \leq z \leq 8$. The distribution of redshifts across the SMACS 0723 field is shown in Figure~\ref{fig:redshifts}. From these 355 sources 280 were considered classifiable, excluding stars and point sources.

As an additional sanity check of these redshifts, and to see if they could be improved upon, we conduct our own SED fitting procedure with the use of LePhare \citep{Arnouts1999,Ilbert2006}. Within LePhare, we use the COSMOS galaxy templates \citep{Ilbert2009} which are based on the commonly used BC03 template set \citep{Bruzual2003}. These templates are modified with dust attenuation up to $E(B-V)=1.5$ \citep{Calzetti2000} and attenuation from the IGM following \citet{madau95}. We initially run the SED fitting process on the original RELICS photometry and obtain strong agreement with their BPz based redshifts. We then add in photometry from an F200W selected NIRCam catalogue which is cross matched to the RELICS catalogue with a 0.5 arcsecond tolerance. To be consistent with the photometry derived in the RELICS catalogues, we use isophotal magnitudes as measured by SExtractor \citep{Bertin1996} for our sources. We find that the photo-z's are consistent with the original RELICS estimations when the bluest NIRCam bands available (F090W and F150W) are added to the SED fitting procedure. However, when photometry from F200W and redwards are added, we find that some originally high redshift sources ($z>3$) in the RELICS catalogue are given new solutions at $z<1$, this is found to be the result of a lack of a strong Balmer break at $\sim2\mu$m. Examining these sources in detail reveal them to be classified as a star-forming disk in our later analysis. Estimations of the proper size and absolute magnitudes of these objects reveal many of the new low-z solutions to give extremely small ($\leq0.1 \ $pkpc) and faint ($M_{F415W}>-16$) properties for the sources\footnote{The results presented here are robust against putting these sources at lower redshifts since they represent a small fraction of the overall sample.}. Additionally, the current dataset was reduced prior to the calibration updates based on on-flight observations, which had large offsets on zero-points of up to $\sim 0.4$ mag \citep{Adams2022,Rigby}. This observation, combined with the subsequent disk classification lead us to proceed with the original RELICS BPz redshifts for all sources. 

\subsection{Quantitative morphologies: \textsc{Morfometryka}}\label{sec:mfmtk}

\begin{figure*}
    \centering\includegraphics[width=0.95\textwidth]{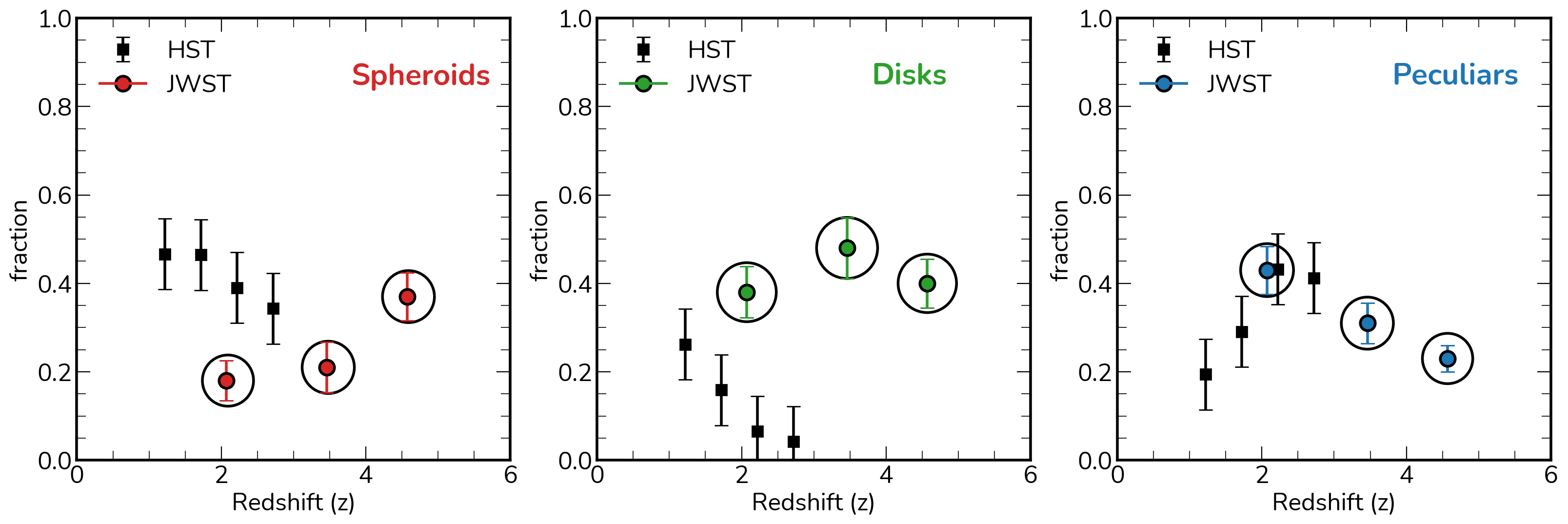}
    \caption{Plots showing the morphological evolution of the galaxies found in the SMACS 0723 field up to $z = 6$.  These show the fraction of the total number of galaxies, within a given redshift bin, which has the given type as determined by visual morphologies.    We also include the morphological evolution which has been derived from HST observations of the CANDELS fields in \citet[][]{Mortlock2013}.  Circled markers denote the JWST observations at higher redshifts. We note that the increase in spheroids can be attributed to smaller sizes with increasing redshift $z$ as discussed in \S~\ref{sec:sph}.}
    \label{fig:mfmtk}
\end{figure*}

\textsc{Morfometryka} was designed to perform several structural measurements on galaxy images, in an automatic non-interactive way \citep{Ferrari2015, Lucatelli2019}. It was devised mainly to measure non-parametric morphometric quantities, but also performs single-component Sérsic model fitting. It takes as input the galaxy and point spread function (PSF) images, estimates the background with an iterative algorithm, deblends the sources and defines which one is the target. Then, it filters out external sources using GalClean\footnote{\href{https://github.com/astroferreira/galclean}{GalClean -- https://github.com/astroferreira/galclean}} \citep{FERREIRA2018}. From the segmented region it calculates basic geometrical  parameters (e.g.~center, position angle, axial ratio) using image moments. Following this, it performs photometry, measuring fluxes in ellipses with the aforementioned parameters. Along the way, it masks point sources over the ellipse annulus with a sigma clipping criterion.

From the luminosity growth curve it establishes the Petrosian radius  and the Petrosian Region, inside which all measurements are made. The 1D Sérsic fit is performed on the luminosity profile. For robustness, the 1D outputs are used as inputs for a 2D Sérsic fit done with the galaxy image and JWST PSF images generated with the official package \texttt{WebbPSF}\footnote{\url{https://webbpsf.readthedocs.io/}}. \textsc{Morfometryka} uses the PSF to produce the Sérsic profiles and to mask an area of the size of the PSF FWHM from the central region of the source stamp for non-parametric morphology calculations. Even though these simulated PSFs are realistic, we note that deviations from the true PSF might exist. However, as we are interested in extended sources, effects of this type are neglible. Finally, \textsc{Morfometryka} measures several morphometric parameters (concentration; asymmetry; Gini; M20; entropy, spirality, and curvature, among others).

\subsection{Visual Classification}\label{sec:visual}

All galaxies in the sample were classified by three co-authors of this paper with experience with galaxy structure and classification (CC, LF, ES). The visual classification scheme that we use is described in detail in Ferreira et al.~(2022; in prep).  Here we give a quick summary. In general, we use four categories for visual classifications, following \citet[][]{Mortlock2013}; these are defined as described below, and differ slightly from a traditional Hubble classification scheme, but are generally very similar. 

\begin{itemize}
\item Class 0: Unclassifiable: Galaxies too small and/or too faint to classify, and images with artifacts. 
\item Class 1: Spheroids: These galaxies are resolved, symmetrically and centrally concentrated, with a smooth profile and are round/elliptical in shape. 
\item Class 2: Disks: This category includes galaxies that exhibit a resolved disk in the form of an outer area of lower surface brightness with a regularly increasing brightness towards the center of the galaxy. This classification does not depend on there being a spiral pattern in the system, although one can be present in this classification. 
\item Class 3: Peculiar. This class is for well-resolved galaxies with a morphology which is dominated by a disturbance or peculiarity and has no obvious disc or spheroid component. 
\end{itemize}


Each galaxy is further classified as smooth or structured, where structured galaxies have features standing out from the smooth stellar envelope, such as star-formation clumps, tidal features, and merger signatures. Galaxies with distinct disk and bulge components were also classified as structured. Finally, the classifiers were able to provide additional notes on each source to aid in future analysis. We do not however use these further detailed morphologies in this paper.

\begin{figure*}
    \centering
    \includegraphics[width=0.95\textwidth]{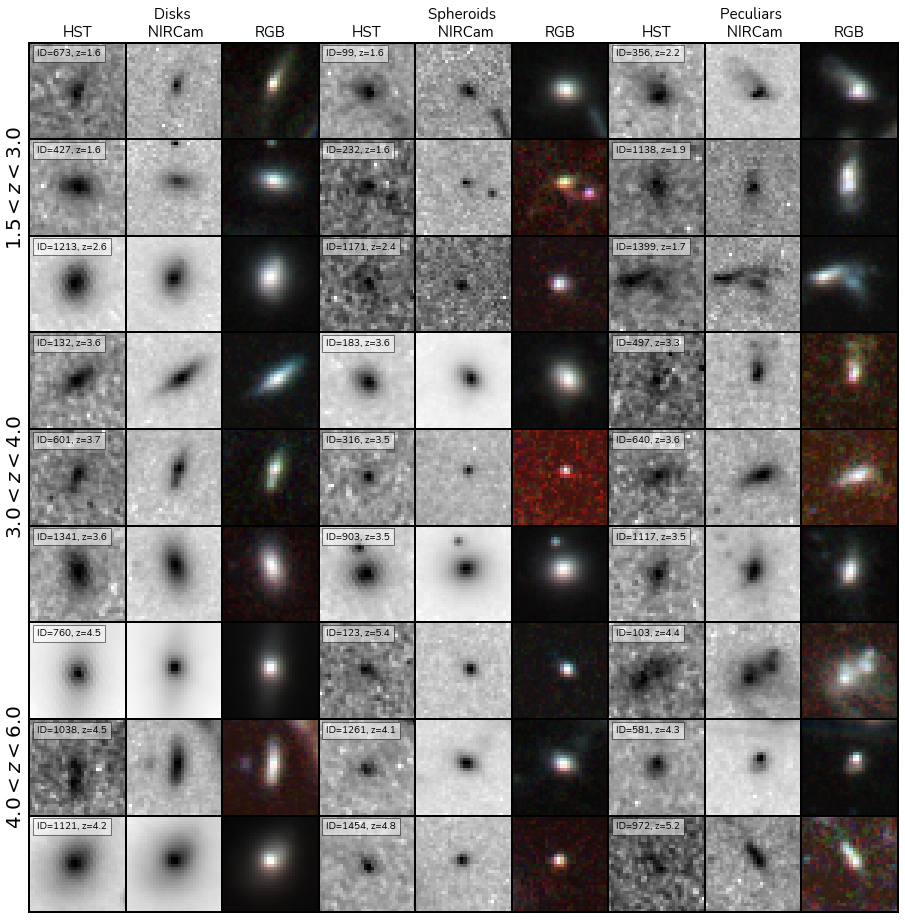}
    \caption{Comparison HST vs JWST images for 9 objects in each class within our sample. Left columns shows the HST F160W image, whilst the middle panel shows the JWST view at the same orientation, in the rest-frame optical.  The far right panel shows the color image of this system as seen through the JWST NIRCam F277W, F356W and F444W filters, generated by \textsc{TRINITY} \citep{Trinity}. The IDs refer to the original RELICs catalogs identification.}
    \label{fig:sersic}
\end{figure*}

These classifications were all carried out separately and then combined into an average, which we then use throughout this work.  Our process was such that we only classified each galaxy in our sample in the wavelength which most closely matches the rest-frame optical wavelength of the observations.  This allows us to match the classifications at different redshifts to determine how morphological evolution is occurring. We find that the classifications by the three classifiers agree $2/3$ in $\sim63\%$ (177) of the sources while perfect agreement $3/3$ happens $\sim33\%$ (87) of the time. Catastrophic classifications where all 3 classifiers disagree happens only in $\sim5\%$ (16) of the cases.  Table~\ref{tab:class} shows the average of the fractions for each of the three classifiers for the three different types (sph, disk and peculiar).  Also listed as the error-bar on these averages is the standard deviation of the fractions among the classifiers, showing that these are always relatively small and in some cases the agreement is to within a few percent.  

\begin{table*}
\caption{The filters used in this study with the redshift ranges used within that filter and the average rest-frame wavelength in which that filter probes at that redshift.  Also shown are the classification fractions for our three main types of galaxies; spheroids, disks, and peculiars.  These are the averages of the three classifiers with the standard deviation listed as the value $\sigma$. Sérsic indices for each class and redshift bin is also provided, showing the mean of the distribution together with $15\%$ and $85\%$ percentile scatter.}
\begin{center}
\begin{tabular}{lccccccccccc}
\hline
Filter & z & $<$Rest-$\lambda>$  & class (sph) & $\sigma$(sph) & class (disk) & $\sigma$(disk) & class (pec)  & $\sigma$(pec) & $n_{sph}$ & $n_{disk}$ & $n_{pec}$  \\
\hline
F090W & 1.5-3.0 & 0.36-0.22 $\mu m$ & 0.18 & 0.02 & 0.38 & 0.10 & 0.43 & 0.12 & $1.32_{1.14}^{1.6}$ & $1.12_{0.58}^{1.62}$ & $0.76_{0.37}^{1.54}$\\
F227W & 3.0-4.0 & 0.69-0.55 $\mu m$ & 0.21 & 0.08 & 0.48 & 0.05 & 0.31 & 0.03 & $1.22_{1.12}^{1.32}$ & $1.04_{0.8}^{1.3}$ & $1.04_{0.31}^{1.53}$ \\
F356W &  4.0-6.0 & 0.71-0.50 $\mu m$ & 0.37 & 0.01 & 0.40 & 0.02 & 0.23 & 0.03 & $1.58_{1.08}^{1.81}$ & $1.11_{0.81}^{1.29}$ & $1.67_{1.26}^{1.09}$ \\
\hline
\end{tabular}
\label{tab:class}
\end{center}
\end{table*}

\section{Results}

\subsection{Distribution of Morphology with Redshift}

One of the main questions that we can investigate with this early imaging from JWST is the distribution of morphological types with redshift.  Given the redshifts we have from HST and the morphologies from JWST we can make the first measurement of the morphological distribution of galaxies up to $z < 8$.  

In Figure~4 we plot the morphological distribution of our sample of galaxies with morphological classifications from the JWST imaging.
As can be seen, we find a remarkable increase in the number of spiral galaxies over what was thought to exist in previous analysis of the deepest HST imaging in the NIR, which found that that there were very few disk galaxies at $z > 1.5$, with a rapid decline in the numbers at higher redshifts  \citep[e.g.,][]{conseliceandA2009, Mortlock2013, Huertas2016}.  

More generally, a decline in spheroids was also seen, but the morphological change with redshift was not as pronounced as it was for the disk galaxies. Figure~5 shows some examples of different galaxy types and how they appear differently in the JWST vs. the HST imaging, revealing that morphologies are often much easier to make out within the JWST data.  There is however, a propensity for these disks and spheroids to contain peculiar features, such as tidal features and clumpy regions, that can differentiate them from $z = 0$ examples.  We however do not investigate these further in this paper.

Overall, we find that the disk galaxy population makes up about half of the galaxies that are identified within the field of SMACS 0723 at $z > 2.5$.  This is a remarkable result, as it shows that galaxies such as the Milky Way could potentially have retained the same overall morphological state for over 12 billion years if these distant disk galaxies are similar to the ancestor galaxy of the Milky Way.

\subsubsection{Morphology at $z > 4$}\label{sec:sph}

The morphologies of $z > 4$ galaxies, as probed by the F356W filter, differ from the rest-frame UV morphologies determined from HST imaging \citep[][]{conseliceandA2009}.  Despite what might have been expected from HST observations, the morphologies of at least the brightest galaxies are much less distorted than had been previously been thought based on HST observations.  Also, the galaxies at this epoch are often very tiny, such that their size in a NIRCam image is just a bit larger, or within,  the PSF of JWST. In fact, the larger number of spheroids/compact objects is in part due to the fact that so many of these systems are unresolved, an indication that their sizes are quite tiny.  Future studies carefully measuring sizes with the use of the JWST PSF will examine these sizes and their evolution.  

\subsection{Quantitative Morphologies}

\begin{figure}
    \centering
    \hspace{-1cm} \includegraphics[width=0.5\textwidth]{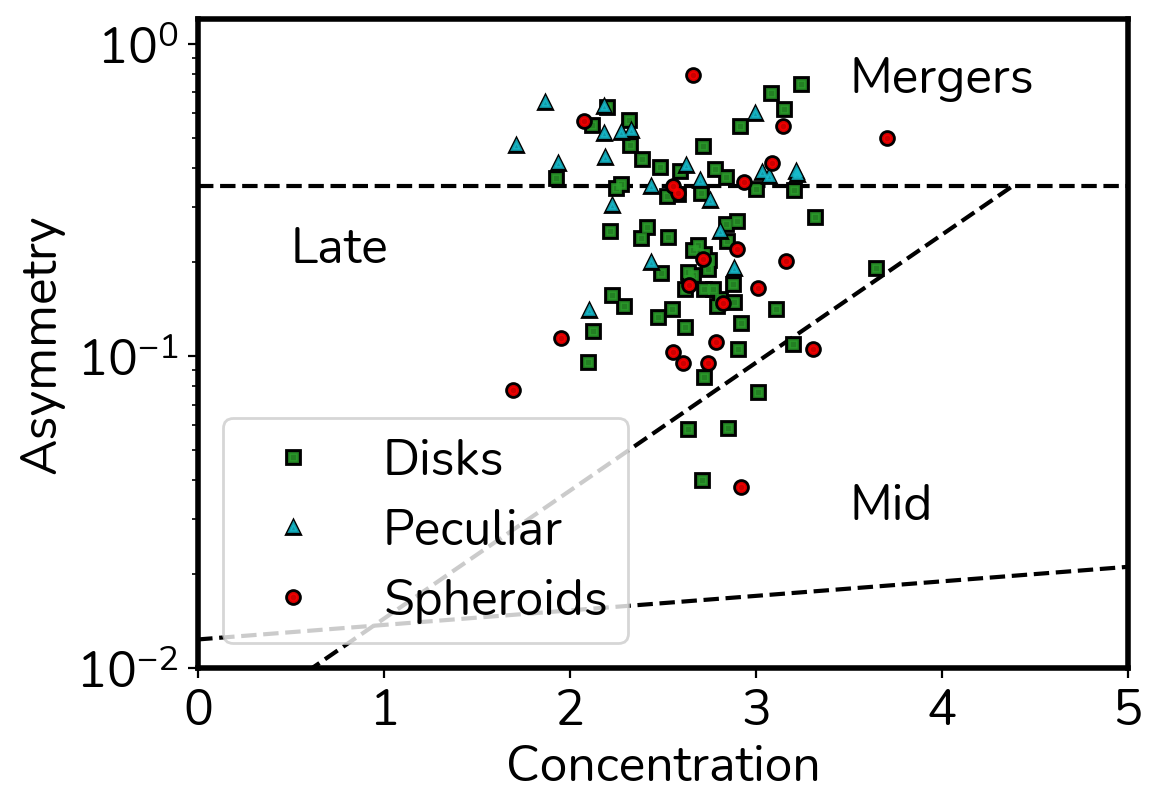}
    \caption{Concentration vs. Asymmetry. The abundance of disks results in their populating the area originally used for selecting late-type galaxies. Decision thresholds shown here are drawn from \citep[][]{Bershady2000}.}
    \label{fig:sersic}
\end{figure}

\begin{figure}
    \centering
    \includegraphics[width=0.45\textwidth]{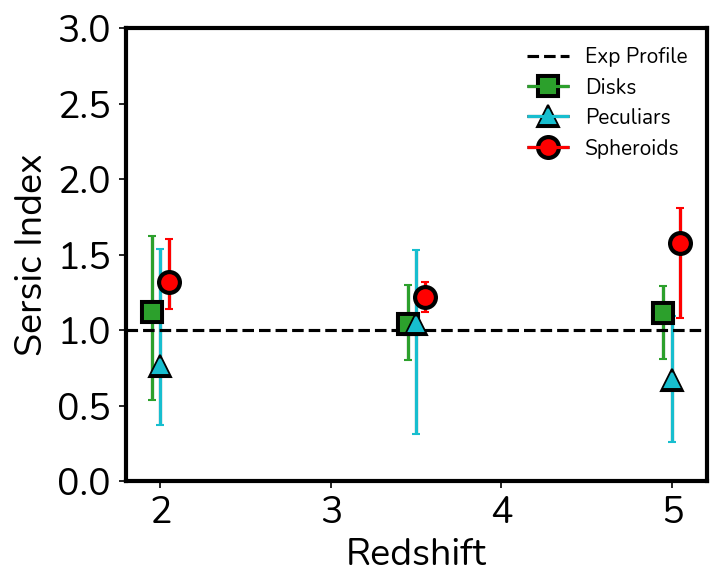}
    \caption{Plot showing the Sérsic index evolution in bins of redshift.  Shown are the morphological types and their mean values for these indices alongside error bars representing the $15\%$ and $85\%$ percentiles of the distribution.  We can see that the average hovers about $n = 1$, but that the spheroids appear to have higher values even at the higher redshifts.}
    \label{fig:sersic}
\end{figure}

We quantify the structures of these galaxies based on the NIRCam imaging. This can be done in a number of ways, and will be the focus of future dedicated papers. We give a broad overview of quantitative morphology for our sample and leave it to future papers to elaborate on these issues.  

First we show the concentration-asymmetry diagram, which has been used to classify galaxies and has a broad correspondence with galaxy types in the nearby universe \citep[e.g.,][]{Conselice2000, Conselice2003, Conselice2008,Bluck2012, Whitney2021}. We present the concentration-asymmetry diagram in the SMACS 0723 field in Figure~6 as measured by \textsc{Morfometryka}.    What we find is that there is no great distinction between the disks and the spheroids, but we do find that the peculiars are in the region of high asymmetry where mergers are located \citep[][]{Conselice2000, Whitney2021}.   We also find that there are few galaxies with very high concentration values, consistent with previous work that found even massive galaxies to have low light concentrations \citep[][]{Buitrago2012}.

Another avenue of investigation is the examination of the light profiles of our galaxies, which we have also measured.  Previous work has shown that almost all massive galaxies at $z > 2$ have Sérsic indices which are $n \sim 1$, which differs for galaxies at lower redshifts where $n \sim 4$ for similar mass galaxies \citep[e.g.,][]{Buitrago2012, Bluck2014}.  We show a basic view of the average Sérsic index evolution for our sample in Figure~\ref{fig:sersic}, which demonstrates that many of our galaxies contain indices with $n\sim1$, with most disks around this value, as expected.  We also find many spheroids at this Sérsic index,  but on average these spheroids have a larger $n$ value.The mean values together with $15\%$ and $85\%$ percentiles of the distributions for each redshift bin and class are displayed in Table~\ref{tab:class}.

\subsection{Formation of the Hubble Sequence}

One of the primary goals of galaxy morphology and structural analysis is to determine when the Hubble sequence was established.   By Hubble sequence we mean the establishment of spheroids (e.g., ellipticals) and spiral galaxies as we see in the case of the most massive galaxies in the nearby universe.  We know for certain that there are fewer ellipticals and spirals at high redshift. However, an important question is: when did the first spheroids and disk galaxies form?  

It is important to be clear about what we mean by this, as a definition of these galaxy types is neither trivial nor simple.  By a `spheroid' we mean a galaxy that exhibits a round or elliptical shape with a classical, steep light profile and a smooth structure. A `disk galaxy' is one that is either a smooth, disk-like object, or something with visible spiral arms. 

The trend of galaxy type with redshift has been measured by several different papers. We include the analysis in \cite{Mortlock2013} as the basis for our understanding of the morphological evolution at $z < 3$.  As Figure~4 shows, we have not reached the limit of where the first ellipticals and spheroids have formed.  We will need to probe even higher redshifts to find when and if there are no spheroids or disk galaxies.  Thus, at least some aspect of the Hubble sequence was in place at $z \sim 6$.  It is, however, important to point out that these classifications are done purely by visual estimates in one band. We have not used colour or other features to classify galaxies, and we know from work with WFC3 that galaxy structure and physical properties of their stars becomes decoupled at higher redshifts $z > 1.5$ \citep[][]{Conselice2011}. It remains to be seen how the physical properties of our `Hubble types' here correlate with the underlying stars in these systems.

\section{Discussion}

This is one of the earliest papers on the morphologies of galaxies at high redshift with JWST, and thus our conclusions will be revisited by others in the months and years to come.  However, it does appear from an initial analysis that there are far more disk galaxies at high redshift than originally thought with HST.  We in fact find that at the highest redshifts probed by HST there are in fact up to 10 times more disk galaxies than we had thought, based on the JWST visual morphologies. 

This implies that disk galaxies have existed in large numbers for quite a significant amount of time.  This may mean that the morphologies of some disk galaxies, such as the Milky Way, have remained in their current form for over 10 billion years.  This would challenge our ideas about mergers being a very common process, and it might be the case that mergers are only a dominant process for forming the stellar masses of certain types of galaxies, namely spheroids, which have a relatively constant merger fraction at $z > 2.5$ at around 10\%.   Although on average galaxies should go through multiple mergers over cosmic time \citep[][]{duncan2019}, it is not clear how these mergers would affect disk morphologies or if there are only certain galaxies that go through mergers multiple times while others, such as the disks we find here, do not undergo these mergers very often or at all at $z < 6$. 

Alternatively, it is also possible that these high redshift disks undergo major mergers, but reform their disks after the disruptive event. This is a process that is found to happen in simulations of gas-rich mergers \citep[e.g.,][]{Sparre2017, Peschken2020}.

There are a few caveats with this study that future studies will be able to flesh out in much more detail. The first is that we only use the visual rest-frame optical morphology of a galaxy to determine whether or not it is a spheroid, disk, or a peculiar.  These systems, however, are more obvious than they were in the HST imaging, implying that in the rest-frame optical we are seeing the underlying morphology in a much clearer way than we are in the rest-frame UV, despite strongly star-forming galaxies having a very similar appearance in the UV and optical, at least at $z < 3$ \citep[][]{Windhorst2002, Papovich2005, Taylor2015, Mager2018}.  It would appear that at least disk galaxies are not easily seen in the UV, and this is an indication that their stellar population and star formation histories are spatially segregated (old/young stars in bulges/disks, for example), just as they are at lower redshifts.  Future studies will certainly be able to study these resolved structures in more detail to learn about the detailed process of disk formation, as done for HST observations at $z < 3$ \citep[][]{Huertas2016, Margalef2022}.

Finally, there is also the fact that this was conducted in a small field of view area of 2.2'x2.2', around a lensing cluster. In the future we will probe non-cluster regions and larger areas to allow for a more detailed comparison with previous results from HST imaging, such as CANDELS \citep{Grogin2011, Anton2011}.

\section{Summary and Conclusions}

In this paper we present a morphological and structural analysis on some of the earliest galaxies imaged by the JWST, which has provided rest-frame optical morphologies and structures for a statistically significant number of galaxies at $z > 3$ for the first time. We also examine the structures of galaxies at $1.5 < z < 3$, where HST has not had the depth and resolution to infer galaxy morphology correctly.   Three of the authors classified 280 galaxies visually at $1.5 < z < 8$ to determine basic morphological types - spheroid, disk, peculiar, at rest-frame optical wavelengths given by JWST.  We also ran quantitative parametric and non-parameteric morphologies on these galaxies.

Our key findings are:

I. The morphological types of galaxies changes less quickly than previously believed, based on precursor HST imaging and results.  That is, these early JWST results suggest that the formation of normal galaxy structure was much earlier than previously thought.

II. A major aspect of this is our discovery that disk galaxies are quite common at $z \sim 3-6$, where they make up $\sim 50$\% of the galaxy population, which is over 10 times as high as what was previously thought to be the case with HST observations.  That is, this epoch is surprisingly full of disk galaxies, which observationally we had not been able to determine before JWST.

III. Distant galaxies at $z > 3$ in the rest-frame optical, despite their appearance in the HST imaging, are not as highly clumpy and asymmetric as once thought.  This effect has not been observed before due to the nature of existing deep imaging with the HST which could probe only ultraviolet light at $z > 3$.   This shows the great power of JWST to probe rest-frame optical where the underlying mass of galaxies can now be traced and measured.

This study is the first examination of the problem of distant galaxy morphology with JWST, and specifically the formation of galaxy structure at $z > 3$. Our results suggest many directions for immediate future study.  We have not included any new JWST galaxies that were not seen with HST, and have not examined the structural properties as a function of stellar mass or other physical properties. All of these will need to be fully examined in the future.  

The present study, however, shows the importance of JWST for understanding the structural evolution of galaxies, which is now open for detailed investigation.

\acknowledgements 

We thank the anonymous referee and Dan Coe for comments that led to improvements in this manuscript. We thank Anthony Holloway, Sotirios Sanidas and Phil Perry for critical and timely help with computer infrastructure that made this work possible.
We acknowledge support from the ERC Advanced Investigator Grant EPOCHS (788113), as well as a studentship from STFC.  This work is based on observations made with the NASA/ESA \textit{Hubble Space Telescope} (HST) and NASA/ESA/CSA \textit{James Webb Space Telescope} (JWST) obtained from the \texttt{Mikulski Archive for Space Telescopes} (\texttt{MAST}) at the \textit{Space Telescope Science Institute} (STScI), which is operated by the Association of Universities for Research in Astronomy, Inc., under NASA contract NAS 5-03127 for JWST, and NAS 5–26555 for HST. These observations are associated with program 14096 for HST, and 2736 for JWST. LF acknowledges financial support from Coordenação de Aperfeiçoamento de Pessoal de Nível Superior - Brazil (CAPES) in the form of a PhD studentship.

\bibliography{corrected_refs}{}
\bibliographystyle{aasjournal}



\end{document}